\documentclass[conference]{IEEEtran}
\usepackage{amsmath}
\usepackage{amsfonts}
\usepackage{amssymb}
\usepackage[pdftex]{graphicx}

\newtheorem{theorem}{Theorem}

\renewcommand{\vec}[1]{\mathbf{#1}}
\newcommand{\defn}[1]{\emph{#1}}

\newcommand{\mat}[1]{\mathsf{#1}}

\newcommand{\K}{\mathcal K}

\newcommand{\I}{\mathbb {I}}

\newcommand{\zero}{\mathtt 0}
\newcommand{\one}{\mathtt 1}
\newcommand{\zeroone}{\{\mathtt 0,\mathtt 1\}}
\newcommand{\NA}{\text{NA}}
\newcommand{\A}{\text{A}}
\renewcommand{\L}{\mathcal L}
\newcommand{\Khat}{\hat{\mathcal K}}

\renewcommand{\I}{I}
\newcommand{\HH}{H}

\newcommand{\blah}[1]{}

\begin{document}

\title{Adaptive Group Testing as \\ Channel Coding with Feedback}

\author{\IEEEauthorblockN{Matthew Aldridge}
\IEEEauthorblockA{Heilbronn Institute of Mathematical Research\\ School of Mathematics\\
University of Bristol, UK\\
 m.aldridge@bristol.ac.uk}}

\maketitle

\begin{abstract}
Group testing is the combinatorial problem of 
identifying the defective items in a population by grouping items into test pools.
Recently, nonadaptive group testing -- where all the test pools
must be decided on at the start -- has been studied from an information
theory point of view. Using techniques from channel coding, upper and lower bounds
have been given on the number of tests required to accurately recover the defective set,
even when the test outcomes can be noisy.

In this paper, we give the first information theoretic result on
adaptive group testing -- where the outcome of previous tests can
influence the makeup of future tests.  We show that adaptive testing does
not help much, as the number of tests required obeys the same lower bound
as nonadaptive testing.
Our proof uses similar techniques to the proof that
feedback does not improve channel capacity.
\end{abstract}


\section{Introduction}

The problem of \defn{group testing} concerns detecting the defective members of a set of items through the means of pooled tests.  Group testing as a subject dates back to the work of Dorfman \cite{dorfman} in 1940s studying practical ways of testing soldiers' blood for syphilis, and has received much attention from combinatorialists and probabilists since.

The setup is as follows: Suppose we have a number of \defn{items}, of which some are \defn{defective}.
To identify the defective items we could test each of the items individually for defectiveness.
However, when the proportion of defective items is small, most of the tests will give negative results.
A less wasteful method is to test \defn{pools} of many items together at the same time.
In the noise-free model, a pool gives a negative test outcome if it contains no defective
items, and gives a positive outcome if it contains at least one defective item.
(In Section II of this paper we consider models with noise.)
After a number $T$ of such pooled tests, it should be possible to deduce which items were defective.

Traditionally, group testing has been seen as a combinatorial problem. 
One aims to find a pooling strategy such that each possible defective set
gives a different sequence of outcomes.  This gives a zero error probability, and one is interested in how small $T$ can be made. (See, for example, the textbook of Du and Hwang \cite{du} for more details on the combinatorial approach to group testing.)

Group testing splits into two main types:
\begin{itemize}
  \item \textbf{Nonadaptive group testing}, where the entire pooling strategy
    is decided on beforehand;
  \item \textbf{Adaptive (or sequential) group testing}, where the outcomes of previous tests
    can be used to influence the makeup of future pools.
\end{itemize}

Recently, new results on nonadaptive group testing with arbitrarily small
probability of error have been derived using
information theoretic techniques. A recent paper of Atia and Saligrama
\cite{atia} proves bounds on $T$ using techniques similar to the proof of
Shannon's channel coding theorem \cite{shannon1}.

In this paper, we study adaptive group testing using information theoretic
techniques.  Clearly adaptive group testing cannot be more difficult than
nonadaptive testing. We show that it is not much easier either.  

Specifically, Theorem 2
shows that the number of tests required for adaptive group testing is no more
than that required for nonadaptive testing, but is still greater than the
Atia--Saligrama lower bound.  The result is obtained by techniques similar to
Shannon's proof that feedback does not improve capacity for channel coding \cite{shannon2}.
As far as we are aware, this is the first
information theoretic result for adaptive group testing.

In combinatorial zero-error group testing using the noise-free model,
adaptive testing certainly is an improvement.
Only $O(K \log N)$ are needed for adaptive testing, whereas at least
$\Omega (K^2 \log N/\log K)$ are required for nonadaptive testing
\cite{du}. We note that this is similar to the case of zero-error
channel coding, where feedback may improve the zero-error capacity
\cite{shannon2}.

The structure of this paper is as follows. In Section II we outline
the information theoretic approach to nonadaptive group testing, fixing notation,
and reviewing the work of Atia and Saligrama \cite{atia} and others.  In Section
III we briefly review Shannon's result on channel coding with feedback
before stating and proving our main theorem (Theorem 2). We conclude with
Section IV.

\section{The information theoretic approach\\ to nonadaptive group testing}

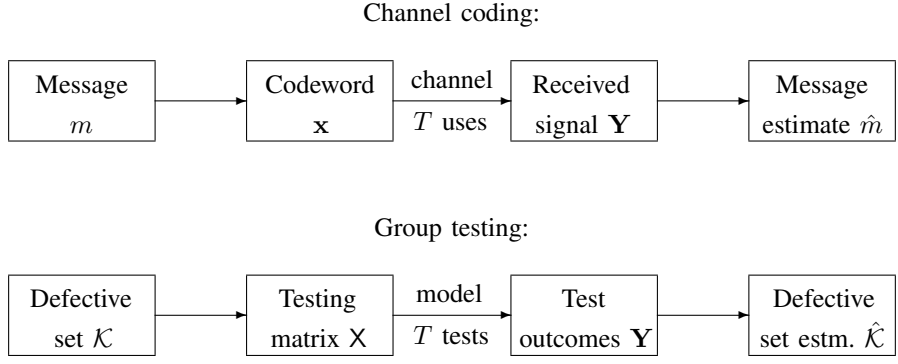
\begin{figure*}
	\begin{center}
	
	Channel coding:

	\smallskip	
	
		\begin{picture}(335,40)(0,0)

		  \put(0,0){\framebox(55,30)}
		  \put(10,18){Message}
		  \put(23,3){$m$}
		  \put(24,33){}
		  
   	      \put(55,15){\vector(1,0){35}}
		  \put(71,18){}
		  
		  \put(90,0){\framebox(55,30)}
		  \put(97,18){Codeword}
		  \put(115,3){$\vec x$}
		  
		  \put(145,15){\vector(1,0){45}}
		  \put(152,20){channel}
		  \put(108,33){}
		  \put(153,4){$T$ uses}
		  
		  \put(190,0){\framebox(55,30)}
		  \put(198,18){Received}
		  \put(199,3){signal $\vec Y$}
		  \put(214,34){}
		  
		  \put(245,15){\vector(1,0){35}}
		  \put(259,18){}
		  
		  \put(280,0){\framebox(55,30)}
		  \put(290,18){Message}
		  \put(285,3){estimate $\hat m$}
		  \put(300,33){}
		  
    \end{picture}

  \bigskip
  
  \bigskip

	Group testing:
	
	\smallskip
	
		\begin{picture}(335,40)(0,0)

		  \put(0,0){\framebox(55,30)}
		  \put(8,18){Defective}
		  \put(17,3){set $\K$}
		  \put(24,33){}
		  
     	  \put(55,15){\vector(1,0){35}}
		  \put(71,18){}
		  
		  \put(90,0){\framebox(55,30)}
		  \put(102,18){Testing}
		  \put(99,3){matrix $\mat X$}
		  
		  \put(145,15){\vector(1,0){45}}
		  \put(154,20){model}
		  \put(108,33){}
		  \put(153,4){$T$ tests}
		  
		  \put(190,0){\framebox(55,30)}
		  \put(209,18){Test}
		  \put(193,3){outcomes $\vec Y$}
		  \put(214,34){}
		  
		  \put(245,15){\vector(1,0){35}}
		  \put(259,18){}
		  
		  \put(280,0){\framebox(55,30)}
		  \put(289,18){Defective}
		  \put(284,3){set estm.\ $\Khat$}
		  \put(300,33){}
		  
    \end{picture}

\end{center}
\caption{A diagram illustrating the similarities between channel coding and group testing.}
\end{figure*}

First we fix some notation.  We have $N$ items, of which a subset $\K$ of size $K$ is defective.
We wish to accurately estimate the defective set from $T$ tests.
A pooling strategy can be defined by a \defn{testing matrix}
$\mat X = (x_{it}) \in \zeroone^{N\times T}$, where $x_{it} = \one$
denotes that item $i$ is in the pool for test $t$, and $x_{it} = \zero$
denotes that it is not.  Test $t$ gives an output $Y_t$ in some
output alphabet $\mathcal Y$ (which is usually $\zeroone$ also). Then, given the test outcomes
$\vec Y = (Y_t) \in \mathcal Y^T$, we make
an estimate $\hat\K = \hat\K(\vec Y)$ of the defective set.
The average probability of error is $\epsilon$.

Let $k_t = |\{i \in \K : x_{it} = \one \}|$
denote the number of defective items in test $t$. In the main \defn{noise-free} case, 
$Y_t = \zero$ (denoting a negative test outcome) if $k_t = 0$, and
$Y_t = \one$ (denoting a positive test outcome) if $k_t \geq 1$.

We can also consider group testing with noise. Atia and Saligrama \cite{atia} consider
two noise models:
\begin{itemize}
  \item \textbf{Addition model}, where false positives occur with
    probability $q$.  That is,
      \begin{align*}
        \text{if $k_t = 0$,}    \qquad Y_t &=
          \begin{cases} \zero & \text{with probability $1-q$,} \\
                        \one  & \text{with probability $q$;} \end{cases} \\
        \text{if $k_t \geq 1$,} \qquad Y_t &= \one .
     \end{align*}
  \item \textbf{Dilution model}, where false negatives occur with
    probability $u^{k_t}$.  That is,
      \begin{align*}
        \text{if $k_t = 0$,}    \qquad Y_t &= \zero ; \\
        \text{if $k_t \geq 1$,} \qquad Y_t &=
          \begin{cases} \zero & \text{with probability $u^{k_t}$,} \\
                        \one  & \text{with probability $1-u^{k_t}$.} \end{cases}
     \end{align*}
\end{itemize}
Sejdinovic and Johnson \cite{sej} considered a model where both addition and
dilution errors can occur.  Aldridge \cite[Chapter 6]{me} considered a class of models
where only defects matter, in that the distribution of $Y_t$ depends only on
$k_t$ (and not on how many nondefective items are in a test pool).

Group testing can be considered as being similar to channel coding.
Here, the defective set takes the place of the message, the testing matrix
is like the codebook, the test outcomes like the received signal.
Then, like channel coding, we want to estimate the message/defective set
using as few channel uses/tests as possible while keeping the error
probability low.  Figure 1 illustrates this.

Atia and Saligrama's main result was the following bounds on the number of tests
required to accurately detect the defective set \cite{atia}.

\begin{theorem}
  Consider a group testing model where only defects matter.
  Let $T_\NA^* = T_\NA^*(N,K,\epsilon)$ be the minimum number of tests necessary
  to identify $K$ defects among $N$ items with error probability at most $\epsilon \neq 0,1$.
  Then
    \[ \underline T + o(\log N) \leq T_\NA^* \leq \overline T + o(\log N) \]
  as $N\to\infty$, where
    \begin{align}
      \overline T &=  \min_p \max_{\L\subset\K}
                    \frac{\log \binom{N-K}{|\L|} \binom{K}{|\L|}}
                         {I (\vec X_{\K\setminus\L} : \vec X_{\L}, Y)} , \label{th1} \\
      \underline T &= \min_p \max_{\L\subset\K}
                    \frac{\log \binom{N-|\L|}{K-|\L|} }
                         {\I (\vec X_{\K\setminus\L} : \vec X_{\L}, Y)} . \label{th2}
    \end{align}
  Here, the $X_i$ are IID $\text{Bern}(p)$, $Y$ is related to $\vec X$ through the channel
  model, and $I$ denotes mutual information. We have used the notation $\vec X_\L := (X_i : i \in \L)$ and similar.
\end{theorem}


Atia and Saligrama \cite{atia} proved the theorem for the noise-free, addition
and dilution models. Aldridge \cite[Chapter 6.4]{me} pointed out that their analysis extends
to any model where only defects matter.  Atia and Saligrama \cite{atia} also extended
their result to the $K=o(N)$ asymptotic regime.

The proof of the upper bound is similar to Gallager's proof \cite{gallager} of
the direct part Shannon's channel coding theorem \cite{shannon1}.  Test pools designed
at random,
with $X_{it} = \one$ with probability $p$ and $X_{it} = \zero$ with probability
$1-p$, IID over $i$
and $t$.  Estimation of the defective set is done on a maximum likelihood
basis, in that $\hat K$ is chosen to maximise the probabilty of the outcome $\vec Y$
given the testing matrix $\mat X$.

The proof of the lower bound resembles the converse part of Shannon's theorem
(see for example \cite[Section 7.9]{cover}), where Fano's inequality bounds the error probability.
Unfortunately, unlike in Shannon's theorem, we are not so lucky that the
upper and lower bounds asymptotically coincide, although they are close up to
a logarithmic factor in $N$.

There has been other recent work on nonadaptive information theoretic group
testing. Sejdinovic and Johnson \cite{sej} gave accurate asymptotic
expressions for $\overline T$ for the noise-free, addition and dilution models.
Cheraghchi et al \cite{cher1} considered group testing when the makeup of the pools
is constrained by a graphical structure.  Numerous authors \cite{sej,porat,cher2,chan} have
used modern decoding algorithms on nonadaptive group testing simulations.

Some similar work has occured in the compressed sensing community; see the
survey of Malyutov \cite{malyutov}.

\section{Adaptive group testing}

In adaptive group testing, the makeup of a testing pool
can depend on the outcomes of earlier tests, so
\[x_{it} = x_{it}(Y_1,\dots,Y_{t-1}). \]

This is similar to channel coding with feedback, where
future inputs to the channel can depend on past outputs.
Shannon proved that (perhaps surprisingly) feedback does
not improve the capacity of a single-user channel \cite{shannon2}.
Since a transmitter could choose not to use the feedback,
it's clear that the capacity with feedback is at least as
high as the capacity without.  However by being more careful
with Fano's inequality in the proof of the converse, it can
apply to the case of feedback also.  See \cite[Section 7.12]{cover}, for example,
for a detailed proof.

Our result proceeds similarly. Due to the non-tightness of the bounds on testing in the nonadaptive case, we will not be able to show that adaptive group testing requires the \emph{same} number of tests as nonadaptive testing, but we will be able to show that it obeys the same lower bound and requires no more tests than the nonadaptive case.

The lack of much improvement due to adaptive testing may initially
seem surprising. However, the analogy with Shannon's feedback result
explains why we should in fact expect this.

We emphasise that our theorem holds not only for the noise-free model, but also
for the dilution and addition models, and any model where
only defects matter.

\begin{theorem}
  Consider a group testing model where only defects matter.
  Let $T^*_{\NA}$ and $T^*_\A$ (dependent on $N$, $K$ and $\epsilon$)
  be the minimum number of tests necessary
  to identify $K$ defects among $N$ items with error probability at most $\epsilon \neq 0,1$ for
  nonadaptive and adaptive group testing respectively.
    Then, as $N\to\infty$, we have the inequalities
    \[ \underline T - o(\log N) \leq T^*_\A \leq T^*_{\NA} \leq \overline T + o(\log N) \]
  where $\underline T$ and $\overline T$ are as in \eqref{th1} and \eqref{th2}.
\end{theorem}


\begin{IEEEproof}
  The third inequality is part of Theorem 1.  The second inequality is trivial, as nonadaptive
  group testing is merely a special case of adaptive group testing where the tester chooses
  to ignore the information of previous test results.
  
  To prove the first inequality, we adapt Atia and Saligrama's proof of
  converse part of Theorem 1 \cite{atia}, and Shannon's proof that feedback fails to improve channel
  capacity \cite{shannon2}, as exposited by Cover and Thomas \cite[Theorem 7.12.1]{cover}.
  
  
  Choose a set of items $\L$ of size $|\L|$ uniformly at random
  from $\{1,2,\dots,N\}$, and choose $\K$ of size $K$ uniformly
  at random from sets containing $\L$.
  
  Suppose a genie reveals to us the $|\L|$ defective items $\mathcal L \subset \K$,
  leaving us to work out the remaining $K-|\mathcal L|$
  defective items.  Given $\mathcal L$, there are
  $\binom{N-|\mathcal L|}{K-|\mathcal L|}$ equally likely choices
  of the random $\K$, so 
    \begin{equation}
      \HH(\K \mid \L) = \log \binom{N-|\L|}{K-|\L|} \label{a} .
    \end{equation}
  Using a standard identity we can rewrite \eqref{a} as
    \begin{equation}
      \log \binom{N-|\L|}{K-|\L|} = \HH(\K \mid \Khat, \L) + \I(\K : \Khat \mid \L)  \label{b} .
    \end{equation}  
  
  We can now use Fano's inequality (see for example \cite[Theorem 2.10.1]{cover})
  to bound the conditional entropy term in \eqref{b} in terms of the error
  probability $\epsilon$. Specifically, we have
    \begin{equation}
       \HH(\K \mid \Khat, \L) \leq 1 + \epsilon \log \binom{N-|\L|}{K-|\L|} , \label{cc}
    \end{equation}
  since there are again $\binom{N-|\L|}{K-|\L|}$ choices for $\K$.
  Substituting \eqref{cc} into \eqref{b} gives
    \begin{equation}
      \log \binom{N-|\L|}{K-|\L|} \leq 1 + \epsilon \log \binom{N-|\L|}{K-|\L|} + \I(\K : \Khat \mid \L) . \label{c}
    \end{equation}

      
  A series of standard information theory inequalities
  and identities show that the mutual information term
  in \eqref{c} can be bounded by
    \begin{equation}
       \I(\K : \Khat \mid \L) \leq T \I (\vec X_{\K\setminus\L} : \vec X_{\L}, Y) . \label{z}
    \end{equation}
  We relegate the elementary (but slightly long-winded)
  verification of \eqref{z} to the Appendix.
    Substituting \eqref{z} into \eqref{c} gives
      \begin{equation}
         \log \binom{N-|\L|}{K-|\L|}
         \leq 1 + \epsilon \log \binom{N-|\L|}{K-|\L|}
         + T \I (\vec X_{\K\setminus\L} : \vec X_{\L}, Y) . \label{alpha}
      \end{equation}
  Rearranging \eqref{alpha} to make $\epsilon$ the subject gives
    \begin{equation}
      \epsilon \geq 1 - T \frac{\I (\vec X_{\K\setminus\L} : \vec X_{\L}, Y)}{\log \binom{N-|\L|}{K-|\L|}}
         - \frac{1}{\log \binom{N-|\L|}{K-|\L|}} . \label{beta}
    \end{equation}
  Sending $N \to \infty$ in \eqref{beta}, it is clear that we require
    \begin{align}
      T &\geq \frac{\log \binom{N-|\L|}{K-|\L|}}{\I (\vec X_{\K\setminus\L} : \vec X_{\L}, Y)}
            \big(1 + o(1) \big) \notag \\
        &= \frac{\log \binom{N-|\L|}{K-|\L|}}{\I (\vec X_{\K\setminus\L} : \vec X_{\L}, Y)} + o(\log N) 
        \label{omega}
    \end{align}
  to force the error probability to be arbitrarily small.
  
  But \eqref{omega} has to be true for all $\L \subset \K$, and we can optimise over the test inclusion
  parameter $p$.  This gives the result.      
\end{IEEEproof}

\section{Conclusion}

In conclusion, we have considered adaptive group testing for
models where only defects matter with arbitrarily low probability of error.
We have shown that adaptive testing requires no more tests than nonadaptive and,
since it still obeys the Atia--Saligrama lower bound, cannot reduce the number
of tests very much.

It remains an open question whether or not $T_\A = T_\NA$ (either
exactly or in an asymptotic sense), or whether, as with zero-error
testing for the noise-free model, there is a gap between $T_\A$
and $T_\NA$.

A `halfway house' between adaptive and nonadaptive testing is \emph{$S$-stage
testing}, where $S$ test pools are decided on at a time.  Clearly
the number of tests required for $S$-stage testing lies between
$T^*_\NA$ and $T^*_\A$ and is nondecreasing in $S$. We are not aware that
this has received any attention from an information theoretic point of view.

\section*{Appendix. \  An inequality about mutual information}

  In this appendix we verify the claim \eqref{z}, that
    \[ \I(\K : \Khat \mid \L) \leq T \I (\vec X_{\K\setminus\L} : \vec X_{\L}, Y) , \]
  which is required in the proof of Theorem 2.
  
  We use the data processing inequality left-hand side of \eqref{z},
  to write
    \begin{equation}
      \I(\K : \Khat \mid \L) 
        \leq \I(\K : \vec Y \mid \L) 
        = \HH(\vec Y \mid \L) - \HH(\vec Y \mid \K) \label{g}
    \end{equation}
  where the second equality in \eqref{g} is standard identity and we have used that
  $\L \cup \K = \K$.
  
  We now unwrap the conditional entropy terms in \eqref{g} using the chain rule
  for entropy (see for example \cite[Theorem 2.5.1]{cover}) and
  standard identities and inequalities. This gives
    \begin{align}
    \begin{split} \I(\K : \Khat \mid \L) 
      &\leq  \sum_{t=1}^T \big( \HH(Y_t \mid Y_1, \dots, Y_{t-1}, \L) \\
        &\qquad\qquad\quad - \HH(Y_t \mid Y_1, \dots, Y_{t-1}, \K) \big)   \end{split} \label{h} \\
        \begin{split}&= \sum_{t=1}^T \big( \HH(Y_t \mid Y_1, \dots, Y_{t-1}, \L, \vec X_{\L t})  \\
            &\qquad\quad - \HH(Y_t \mid Y_1, \dots, Y_{t-1}, \K, \vec X_{\K t}) \big) \end{split}
             \label{i}  \\
        \begin{split}&\leq \sum_{t=1}^T \big( \HH(Y_t \mid \vec X_{\L t}) \\
            &\qquad\quad - \HH(Y_t \mid Y_1, \dots, Y_{t-1}, \K, \vec X_{\K t}) \big) \end{split}
            \label{j}  \\     
        &= \sum_{t=1}^T \big( \HH(Y_t \mid \vec X_{\L t} )
             - \HH(Y_t \mid \vec X_{\K t}) \big) , \label{kay}
    \end{align}
  where we have used the notation $\vec X_{\L t} := (X_{it} : i \in \L)$ for
  fixed $t$ and similar.
  We justify the above steps as follows:
    \begin{itemize}
      \item[\eqref{h}] is from applying the chain rule to the right hand side of \eqref{g};
      \item[\eqref{i}] is because $\vec X_{\L t}$ is a function of
         $Y_1, \dots, Y_{t-1}$ and $\L$, and the same for $\K$;
      \item[\eqref{j}] is because conditioning reduces entropy, so removing conditioning
       increases it;
      \item[\eqref{kay}] is because, conditional on $\vec X_{\K t}$, we
  know $Y_t$ is independent of the previous outcomes $Y_1, \dots,Y_{t-1}$
  and the defective set $\K$.
    \end{itemize}
  
  But the term in the summand of \eqref{kay} is precisely the mutual information
    \begin{equation}
      \HH(Y_t \mid \vec X_{\L t}) - \HH(Y_t \mid \vec X_{\K t})
        = \I (\vec X_{\K\setminus\L\  t} :  Y_t \mid \vec X_{\L t}), \label{aleph}
     \end{equation}
   and this is independent of $t$. Hence substituting \eqref{aleph} into \eqref{kay}
   gives
   \newpage
   \begin{align}
       \I(\K : \Khat \mid \L) &\leq
          \sum_{t=1}^T \I (\vec X_{\K\setminus\L\  t} :  Y_t \mid \vec X_{\L t}) \notag \\
         &= T \I (\vec X_{\K\setminus\L} :  Y \mid \vec X_{\L}) . \label{beth}
     \end{align}

   The mutual information term in \eqref{beth} can alternatively be
  written as
   \begin{align}
      \I(\vec X_{\K\setminus\L} : Y \mid \vec X_{\L})
        &= \I (\vec X_{\K\setminus\L} : \vec X_{\L}, Y) - \I (\vec X_{\K\setminus\L} : \vec X_{\L})
              \notag \\
        &= \I (\vec X_{\K\setminus\L} : \vec X_{\L}, Y) , \label{v}
    \end{align}
   since $\vec X_{\K\setminus\L}$ and $\vec X_\L$ are independent.
   
   Substituting \eqref{v} into \eqref{beth} gives
       \[ \I(\K : \Khat \mid \L) \leq T \I (\vec X_{\K\setminus\L} : \vec X_{\L}, Y) , \]
   thus verifying the claim \eqref{z}.
    
%
%

\section*{Acknowledgments}

The author was supported by the Heilbronn Institute for Mathematical Research
and by the Engineering and Physical
Sciences Research Council, via the University of Bristol 'Bridging the Gaps'
cross-disciplinary feasibility account (EP/H024786/1).
The author thanks Oliver Johnson and Dino Sejdinovic for helpful discussions.




\end{document}